\documentstyle[12pt]{article}
\begin{document}
\title{A Study of  RHIC Crystal Collimation}
\author{\underline{D.Trbojevic}, {\small BNL, USA},
V.Biryukov, {\small IHEP, Protvino, Russia},\\
M.Harrison, B.Parker, P.Thompson, and A. Stevens, {\small BNL, USA},\\
N.Mokhov and A.Drozhdin, {\small FNAL, USA}}
\date{Presented at EPAC 1998 (Stockholm)}
\maketitle
\begin{abstract}
The Relativistic Heavy Ion Collider (RHIC) will experience increasing
longitudinal and transverse heavy ion emittances, mostly due to
intra-beam scattering (IBS). The experiments in
RHIC are expected to not only have reduced luminosities due to IBS
but also an unwanted beam halo.  Primary betatron collimators will
be used to remove the large amplitude particles. The efficiency of
the primary collimator in RHIC depends very much on the alignment
of the jaws which needs to be within few micro-radians for the
best conditions.   As proposed
by V. Biryukov \cite{BiryukovFNAL} bent crystals could be used to improve
the efficiency of an existing collimation system by installing them
upstream of the collimator jaws. Bent crystals have been
successfully used in SPS, Protvino and Fermilab for extraction of
the beam particles channeled through them.  This study examines possible
improvements of the primary collimator system for heavy ions at RHIC
by use of bent crystals.  Bent
crystals will reduce the collimator jaws alignment requirement and
will increase collimator efficiency thereby reducing detector background.
\footnote{Work performed under the auspices of the U.S. Department of Energy}
\end{abstract}

\section{Introduction}
The Relativistic Heavy Ion Collider (RHIC) has
two large angle acceptance detectors PHENIX and STAR placed at the
interaction regions (IR-'s) where the heavy ion beams (like fully
striped gold $^{+79}Au^{197}$ ions) from the two parallel rings
will be focussed to $\beta^{\star}=1-2 m$.
The intra-beam scattering of the heavy ions like
gold $^{+79}Au^{197}$ ions is expected to be a dominant effect
of the emittance growth with a fast loss of luminosity. The transverse
and longitudinal emittances are expected to double in size between
one to two hours due to intra-beam scattering which may lead to
transverse beam loss.  Particle amplitudes also grow due to
other effects like beam gas interaction, beam diffusion due to the
nonlinear beam dynamics etc. This results in beam loss
at limiting apertures at the triplet magnets close to the large detectors.
This beam loss creates hadronic showers which leads to larger than
desirable backgrounds in detectors.
To reduce this background it is necessary to scrape the
unwanted beam.  The primary betatron collimator has to be able to
remove particles with large amplitudes.
The RHIC collimator studies
previously reported \cite{Trbojevic}, \cite{Trbojevic2},
had shown that a combination of the primary with the
secondary collimators reduced the background around the large detectors.
The studies have shown that losses of the out-scattered ions from
the primary collimators were, as expected, localized at the high beta
focusing quadrupoles. The amplitude growth of the
the gold $^{+79}Au^{197}$ and proton ion beams was estimated by a
diffusion process  based on measurements at the SPS-CERN \cite{Fischer}.
The energy and phase distributions
of the out-scattered ions from the primary collimators were obtained by
using the ``ELSHIM'' Van Ginneken code \cite{Ginneken} , \cite{Ginneken2},
which simulated transport of the 100 GeV gold $^{+79}Au^{197}$ ion beam
through the primary collimator jaws. The previous
study \cite{Trbojevic2} has shown that for the best
collimation of the gold ions the alignment of the primary collimators'
jaws has to be within about 10 $\mu$rad.  This is a report where a simulation
of the bent crystal deflection for the RHIC beam collimation of the heavy
ion beam is described.

There are few motivations for this study and an eventual implementation
of the bent crystals for the collimation purposes:
\begin{itemize}
\item The bent crystal, with a length of the order of 1 cm removes a
request for the high accuracy in alignment of the primary collimator jaws
\item The primary collimators, placed $\sim$7 m downstream of the crystal,
become effectively the secondary collimators, because most of the large
amplitude particles are bent into it.
\item The background of the large detectors will be reduced because
the losses at the high focusing quadrupoles are expected to be reduced.
\end{itemize}

\subsection{Bent Single Crystal Deflection}
The protons and heavy ion (like fully stripped lead $^{208}Pb^{82+}$ ions)
bent single crystal extraction has been demonstrated and
continually used at many laboratories like (SPS, Fermilab, IHEP, Aarhus,
ect.) and previously reported in many articles \cite{Arduini}, \cite{Biino},
\cite{BiryukovPhysRev} ect.  The high energy relativistic particles are
channeled within the bent crystal planes and bent away from the center of
the beam.

The theoretical predictions and experimental results of the crystal
extraction and deflection are very well understood.  The crystal
extraction efficiency is defined as the number of ions extracted by the
crystal with respect to the number of lost particles due to the crystal.
 The theoretical
predictions of the crystal extraction {\cal F} are obtained by
tracking the particles through the potential of the bent
crystal lattice taking into account the multipole and catastrophic
scattering, nuclear and electronic collisions with the crystal
discontinuities, surface effects and dislocations
\cite{BiryukovPhysRev}.
The heavy ions (like $^{+79}Au^{197}$) interact with the
crystal matter differently than protons due to their large mass
and high charge state.  The critical angle
$\psi_{c}$, one of the most important parameters for the
channeling acceptance \cite{Lindhard} scales with the mass and charge as
$\psi_{c} \sim \sqrt{\frac{Z}{p}}$.

\section{Bent Crystal Collimation Simulation in RHIC}

The crystal collimation is clearly a multi turn process. The
multi turn efficiency for the TEVATRON is estimated well above 90$\%$ at the
peak according to our simulation.  In this report, instead of a proper simulation of
the multi turn deflection a
simulation a single pass of the diffused gold ions is performed.
The initial horizontal and vertical phase space distributions
of 19528 gold $^{+79}Au^{197}$ ions in front of the crystal
were obtained by selecting ions which had their amplitudes
grown above 5$\sigma_{x}$ horizontally by diffusion process.
The diffusion was described by the published experimental
and theoretical results from the SPS (CERN) \cite{Fischer}.

All particles in the initial horizontal phase space distribution have
the horizontal distance larger than $x \geq$0.04290 m which
represents 5$\sigma_{x}$ of beam with the $40\:\pi\: mm mrad$ emittance.
The betatron functions at this position of the crystal are
 presented in
Table~\ref{MOP37C-t1}.
Before the gold ions were transported through the 1 mrad bent
silicon single crystal Si(110),
an optimum value for the
crystal length was selected to be {\em l}=1 cm  along the beam direction.
This length provides large enough deflection efficiency and
is still practical for the experimental use.
\begin{table}[htb]
\begin{center}
\caption{Betatron Functions at the Crystal}
\begin{tabular}{|l|c|c|c|c|}
\hline
$\beta_{x}$ (m)  &  $\alpha_{x}$    & $\beta_{y}$ & $\alpha_{y}$ & $D_{x}$ \\ \hline
1222.631         & 29.349           & 365.902     & 13.885       & -1.055  \\ \hline
\end{tabular}
\label{MOP37C-t1}
\end{center}
\end{table}
The gold $^{+79}Au^{197}$ ions, with an input energy
of $\gamma $=108.4, are transported through the lattice of the
bent silicon crystal by the Monte Carlo program ``CATCH'' \cite{Biryukov3}.
The total number of the gold ions which passed the crystal was 18296.
The number of particles which interacted with the crystal nuclei was 5\%.
The simulation of the bent crystal collimation is continued by using the
ions obtained from the output at the crystal edge. Four coordinates
of each ion  (x, x', y, y', dp/p) were used as an input into the tracking
program TEAPOT \cite{Talman}.  All measured systematic
as well as the random multipole errors
in the magnetic field for each magnet were introduced into the lattice as well
as the measured magnet misalignment errors. The bent single
crystal is placed in the lattice at the position downstream of the high
focusing quadrupoles at the IR where the PHENIX detector is located.
The tracking was performed at the top energy of 100
GeV/nucleon for gold for 256 turns.
The rms values for misalignment
of the arc quadrupoles were $\Delta$x,y=0.5 mm
$\Delta \theta$=0.5 mrad, while from the measurements of the triplet
quadrupoles the roll and misalignment errors for the rms values
were $\Delta \theta$=0.5 mrad and $\Delta$x,y=0.5 mm.
During tracking the
RF voltage was included and the longitudinal motion of the surviving
particles could be monitored. Particles which survived, with momentum
offsets within the bucket size limit, executed synchrotron oscillations.
In the phase space of the gold ions at the end of
the crystal, there
are large number of particles with momenta larger than the bucket
size. If these particles are not removed by the primary collimators they
will be lost due to their large momentum offsets.
The crystal length, together with its bending angle, can be further
reduced to affect the momentum loss distribution which will be addressed
in our future simulations.

\subsection{Transverse Phase Space Distribution at the Crystal End}
The position of the primary collimator is 7.68 m downstream of the single crystal.
Each tracking run was performed
with a different position of the primary collimator selected to a distance from
center of the beam, starting with a position completely away from
the beam.
In gold ion
distributions in the horizontal phase space at the downstream edge of the crystal,
there are several important features to
point out:
\newline -The channeled particles are contained in the upper part of the
{\em x} and {\em x'} plot with a slope of $ x'_{ch} \sim $0.073 mrad.
The slope of the gold ions, used for the input into the bent crystal,
is within the range of
$-1.080 mrad< x'_{input} <-1.058 mrad$. A difference between the
average slopes of $ x'_{ch}-x'_{input} \simeq $1 mrad,
shows that the channeled particles are bent for $ \simeq $1 mrad.
\newline
-The group of particles at the lower part of the {\em x} and {\em x'}
plot  with a value of x$\simeq $0.04289 m and with
the slope of {\em x'}$\geq $-1.15 mrad is not channeled. The slope of the
horizontal amplitude {\em x'}~1.15 mrad reaches negative values
larger than the end within the initial distribution (-1.108 mrad).
\newline
-Particles with the slope of the horizontal amplitude between
the channeld ones and the edge clearly interacted with the crystal
lattice.
When the primary collimator was completely away from the beam, the particles
from the crystal are lost mostly at
the first quadrupole downstream of it, as well as at the high focusing
quadrupoles around the IR's where the $\beta^{\star}$=1 m.
In the gold ion
distribution in the horizontal phase space, at the 7.68  m downtream
of the crystal,
particles which
reached the primary collimator in the first pass are presented within a
very narrow region between values of the slope of the horizontal amplitude
$-$1.17 mrad$\leq x'\leq -$0.47 mrad. Particles within the ellipse
survived many turns and reach every turn the primary collimator again.
If all the particles are channeled then the primary collimator could be set
up at the distance close to 0.042 m. It is clear that
 not all the particles are channeled and there will be
losses around the ring even when the primary collimator is set to a
horizontal distance of 5$\sigma_{x}$ of the beam. This is due to the paticles
which come from the crystal with the slopes larger than 1.08 mrad from the
crystal edge.
\section{A Comparison of the Crystal Collimation with the Previous Results}
The large amplitude particles are mostly likely to get lost at
the most limited appertures in the two RHIC accelerator rings which are
located at the high focusing triplet magnets.
The background of the two large detectors will be enhanced by the showers
generated by these lost particles. Table~\ref{MOP37C-t2} represents a
\begin{table}[htb]
\begin{center}
\caption{Ring Losses for Standard Collimation}
\begin{tabular}{|l|c|c|}
\hline
Location & Sec. Opened & Sec. Col. 6$\sigma$ \\ \hline
S.Col.   &  0.00 $\%$  & 62.30 $\%$ \\
q2o8     & 42.38 $\%$  & 16.99 $\%$ \\
q3o8	 &  8.20 $\%$  & 3.30  $\%$ \\
q3i6	 & 14.26 $\%$  & 0.30  $\%$ \\
Pr.Col   & 12.11 $\%$  & 5.10  $\%$ \\ \hline
\end{tabular}
\label{MOP37C-t2}
\end{center}
\end{table}
the losses around the RHIC rings from the particles scattered of the
standard collimation case by using a standard primary and secondary
collimators while table~\ref{MOP37C-t3} shows the losses from
the particles deflected by the crystal.
\begin{table}[htb]
\begin{center}
\caption{Ring Losses for Crystal Collimation}
\begin{tabular}{|l|c|c|}
\hline
Location & Pr. Col. Open & Pr. Col. 5$\sigma$ \\ \hline
Pr.Coll. &  0.00 $\%$    & 99.43  $\%$ \\
q2o8     &  0.70 $\%$    &  0.00  $\%$ \\
q3o8	 &  1.87 $\%$    &  0.00  $\%$ \\
q3i6	 & 14.56 $\%$    &  0.52  $\%$ \\
Crystal  & 2.66  $\%$    &  0.00  $\%$ \\ \hline
\end{tabular}
\label{MOP37C-t3}
\end{center}
\end{table}
In the table~\ref{MOP37C-t3} two
extreme cases are presented when the primary collimator downstream of the
crystal is wide open and when it is set to 5$\sigma_{x}$, the same horizontal
distance as a front edge of the crystal.

\section{Conclusions}
We presented a simulation study where the bent single crystals are used in the RHIC
collimation system. The preliminary results are very encouraging. The 100 GeV/nucleon
gold ions were transported through the bent single crystal Si(110) and the gold ions
from the crystal were tracked through the RHIC storage lattice. A comparison to the
previous results with a standard two stage collimator system shows lower losses
around the ring from the outscattered particles. This would possibly
reduced background at the two large detectors due to smaller losses
at the high focusing limited apperture quadrupoles located around the detectors. The
study of the crystal collimation in RHIC will be concluded by the multiturn
crystal deflection simulation and with a shorter length and smaller bending angle
single crystal. We expect improvement in the momentum distribution of the ions
passed through the crystal and a reduction of lost particles around the ring.

\end{document}